# Channel Coherence Classification with Frame-Shifting in Massive MIMO Systems

Ahmad Abboud[1], Oussama Habachi[1]*, Ali Jaber[2], Jean-Pierre Cances[1] and Vahid Meghdadi[1]

[1] XLIM, University of Limoges, Limoges, France
[2] Department of Statistics, Lebanese University, Nabatieh, Lebanon
{oussama.habachi,ahmad.abboud,jeanpierre.cances,vahid.meghdadi}@xlim.fr

**Abstract.** This paper considers the uplink pilot overhead in a time division duplexing (TDD) massive Multiple Input Multiple Output (MIMO) mobile systems. A common scenario of conventional massive MIMO systems is a Base Station (BS) serving all user terminals (UTs) in the cell with the same TDD frame format that fits the coherence interval of the worst-case scenario of user mobility (e.g. a moving train with velocity 300 Km/s). Furthermore, the BS have to estimate all the channels each time-slot for all users even for those with long coherence intervals. In fact, within the same cell, sensors or pedestrian with low mobility UTs (e.g. moving 1.38 m/s) share the same short TDD frame and thus are obliged to upload their pilots each time-slot. The channel coherence interval of the pedestrian UTs with a carrier frequency of 1.9 GHz can be as long as 60 times that of the train passenger users. In other words, conventional techniques waste 59-uploaded pilot sequences for channel estimation. In this paper, we are aware of the resources waste due to various coherence intervals among different user mobility. We classify users based on their coherence interval length, and we propose to skip uploading pilots of UTs with large coherence intervals. Then, we shift frames with the same pilot reused sequence toward an empty pilot time-slot. Simulation results had proved that the proposed technique overcome the performance of conventional massive MIMO systems in both energy and spectral efficiency.

## 1    Introduction

MASSIVE Multiple Input Multiple Output (MIMO) systems is a promising technology to meet future demands of 5G wireless networks. Indeed, this interesting technology boosts both the spectral efficiency and the energy efficiency [1-2]. In this regard, a large number of antennas at the base station (BS) are utilized to communicate with a significantly smaller number of single-antenna user terminals (UTs) over the same frequency and time domain [3]. However, several challenging problems for massive MIMO with a large number of antennas at the BS persist such as the uplink pilot contamination in Time Division Duplexing (TDD) [1]. It is worth noting that channel state information (CSI) are required at the BSs for multi-user MIMO in order to separate the received uplink signals and to direct each downlink signal towards its desired receiver. Pilot sequences sent by the UTs are used to obtain CSI by estimating the channel responses from the received signals. In fact, during a limited coherence interval and under a TDD reciprocity scheme, UTs in each cell must upload orthogonal pilot sequence to their BS for channel estimation lagged with the uplink data. Then, the BS precede the downlink data based on the estimated channel matrix using uploaded pilots. Indeed, the BS require a very low interference in the pilot transmission phase in order to be able to estimate CSI accurately, which make pilot sequences precious resources for massive MIMO systems. Nowadays, the number of the orthogonal pilots is not problematic since the number of orthogonal pilots is much higher than the number of active UTs per cell. However, in massive MIMO systems, the number of active UTs is expected to increase as possible in order to achieve a high sum spectral efficiency. Therefore, this scenario is fundamentally limited by the number of orthogonal uplink pilot sequences that can be generated, which will lead to pilot reuse in adjacent cells and inter-cell interference, i.e. pilot contamination [4]. In view of this, many research efforts have been made to mitigate uplink pilot contamination in massive MIMO systems [5-11]. Most efforts in this field can be categorized into two groups, pilot-based estimation approach and subspace-based estimation approach [12]. Selective uplink training based on channel temporal correlation had been recently proposed in [13] to reduce training overhead. The authors exploit the temporal correlation of UT channels to classify them into two groups. In each channel block, the BS select part of the UTs for training, while the other UT CSIs are determined by prediction. This approach adds channel prediction complexity to the training channel block, where our proposed technique makes benefit from pilot sparsity to shift frames into empty pilot spectral space.

Note also that 5G will be vital for the continued development of Internet of Thing (IoT) letting devices to communicate over vast distances while reducing latency issues. In fact, IoT represents an umbrella keyword that extends the Internet and the Web into the physical world thanks to the widespread deployment of embedded devices having sensing and communication capabilities. It is obvious that the effects of the IoT will be visible in the next few years in both working and domestic fields. Indeed, a big leap forward is coming, related to the use of the 5G network to connect machines and smart communicating objects. The central issues are how to achieve full interoperability between high mobility UTs and low mobility or static sensors, and how to provide them access with a high degree of smartness. Another crucial problem in enabling large numbers of low-cost sensors is energy efficiency. Indeed, the energy efficiency has received a considerable attention when designing communication protocols. In fact, ecological concerns increasingly attract attention in communication systems. Therefore, novel solutions that maintain a limited energy consumption are always welcomed in both system and device levels. This paper proposes to use heterogeneous TDD frame structure by skipping pilots upload in some frames, for UTs with low mobility such as sensors, to reduce pilot contamination and increase energy efficiency (EE) on the uplink of massive MIMO systems. We introduce a sparse pilot model with time-shifted TDD frames to increase system spectral efficiency (SE) and reuse the same pilot sequence while preserving pilot orthogonality. We classify users into different classes according to their coherence intervals and assign to each class a pool of sparse shifted pilot sequences that best matches their coherence intervals length, taking into consideration the overall system performance. We then prove the effect of our proposed technique through simulation results. Our work is different from [6], [5], [13-15] in the sense that these works present time-shifted pilots in an inter-cell manner. In the aforementioned works, each group of cells uses a time-shift frame, while in our work we group users according to their coherence interval length and we use time shift of unit length equal to an entire time-slot. Furthermore, we skip pilot upload in some time-slots for classes that encountered longer coherence intervals. This skipping makes it possible to reuse the same pilot in the same cell several times without encountering pilot contamination. To the best of our knowledge, UTs classification based on their coherence interval length coped with time-shifted TDD frame structure was not proposed in the literature.

The main contributions of this paper are threefold:

- UTs classification based on their coherence interval length,
- UTs with large coherence intervals will skip uploading pilots for some time slots depending on their assigned classes,
- We shift frames with the same reused pilot sequence toward an empty pilot time-slot to take profit from pilot skipping of other users,
- Skipping sending pilot for some time-slots not only enhances the energy efficiency of the UT since it remains idle during the pilot duration but also increase the number of users that can be served by the BS without any pilot contamination.

The remainder of the paper is organized as follows. The next section describes the system model. In Section III, we discuss the proposed time-shifted pilots and sparse pilot. Section IV is devoted to introduce the classification criteria and UTs classification algorithm. In Section V, we discuss the spectral and energy efficiency. Before concluding the paper in Section VII, we introduce, Section VI, some numerical results to illustrate the performance of our proposed technique.

Notations: In this article, transpose and Hermitian transpose are denoted by $(.)^{\mathrm{T}}$, $(.)^{\mathrm{H}}$, respectively. $(.)^*$ denote the conjugate, $\det(A)$ denote the determinant of A, $\odot$ denote element-wise multiplication and denote by $\|A\|$ the Frobenius norm.

## 2 System Model

We consider a hexagonal cellular system with L cells, assigned with an index in the set $\mathcal{L} = \{1, \dots, L\}$, each served by one BS holding M-antennas and communicating with K single-antenna UTs that share the same bandwidth. Note that the frame structure should be matched to the coherence interval of the UTs so that the channel between them and the BS can be described by a constant channel response within a frame. Specifically, $\tau$ symbols of the frame are allocated for pilot signaling and the remaining $T - \tau$ symbols are used for uplink transmission. According to the conventional massive MIMO system, in order to avoid pilot contamination, $\tau$ pilot symbols can generate $\tau$ orthogonal pilot sequences, and then at most τ UTs can transmit pilots. We assume the use of Orthogonal Frequency Division Multiplexing (OFDM) flat fading channel for each subcarrier. We denote by $g_{jmk} \triangleq [\boldsymbol{G_j}]_{m,k}$ the channel coefficient between the m-th antenna of the j-th BS and the k-th user of the l-th cell.

$$g_{jmk} = h_{lmk}\sqrt{\beta_{lk}}$$

$m$=1,2,…,$M$ ; $k$=1,2,…,$K$ and $l$=1,2,…,$L$

$h_{lmk}$ and $\beta_{lk}$ represent the small-scale fading and large-scale fading coefficients respectively, where $h_{lmk} \sim CN(0,1)$ is statistically independent across UTs and $\sqrt{\beta_{lk}}$ models the geometric attenuation and shadow fading. In addition, $\beta_{lk}$ is assumed to be independent over $m$ and to be constant over many coherence time intervals and known prior.

In general form, we can represent the channel matrix:

$$\boldsymbol{G} = \boldsymbol{H}\boldsymbol{D}^{1/2} \qquad (1)$$

where $\boldsymbol{H}$ is the M×K matrix of fast fading coefficients between the K users and the M antennas of the BS, i.e. $h_{mk} \triangleq [\boldsymbol{H}]_{m,k}$ and D is the K×K diagonal matrix, where: $[\boldsymbol{D}]_{k,k} = \beta_k$ represents the large-scale fading between the BS and the K UTs.

At the reverse-link, the received signal vector of dimension M ×1 at the j-th BS during uplink session can be represented as follows:

$$\mathbf{Y}_j^{up} = \sqrt{P_u}\sum_{l=1}^{L} \boldsymbol{H}_l \mathbf{D}_l^{1/2}\boldsymbol{X}_l^{up} + \mathbf{W}_j^{up} \quad (2)$$

$$\mathbf{Y}_j^{up} = \underbrace{\sqrt{P_u}\boldsymbol{H}_j\mathbf{D}_j^{1/2}\boldsymbol{X}_j^{up}}_{prefered\ signal} + \underbrace{\sqrt{P_u}\sum_{l=1,l\neq j}^{L} \boldsymbol{H}_l\mathbf{D}_l^{1/2}\boldsymbol{X}_l^{up}}_{contaminated\ signal} + \underbrace{\mathbf{W}_j^{up}}_{noise\ vector} \qquad (3)$$

$\boldsymbol{X}_l^{up}$ denotes the K×1 symbol vector uploaded from users in the l-th cell and $\mathbf{W}_j^{up}$ denotes the additive AWGN i.i.d noise vector with zero-mean, unit-variance $CN(0,1)$ received at BS antennas as a vector of dimension M×1. $P_u$ denotes the uplink transmission power of each UT.

We follow the system model represented by [16], and we denote by K' the number of UTs that uploads their pilots in the current time-slot and by K" the number of UTs that does not upload their pilots during the current time-slot, i.e. K= K' +K". Considering the uplink pilot session of length $\tau \times 1$, then the received signal at the j-th BS is expressed as follows:

$$\mathbf{Y}_j^p = \sqrt{\tau P_u}\sum_{l=1}^{L} \boldsymbol{G}_l[\boldsymbol{X}_l^p \odot \boldsymbol{S}_l] + \mathbf{W}_j^{up} \qquad (4)$$

where $\boldsymbol{S}_l$ is a $K \times 1$ binary matrix with elements $s_k \in \{0,1\}$. $s_k = 0$ for UT that uploads a TDD format without pilots and $s_k = 1$ if UT uploads TDD format using pilots. $\boldsymbol{X}_l^p$ is the $K \times 1$ matrix with elements $x_{lk}^p$, where each represents an orthogonal pilot sequence uploaded from the k-th UT of the l-th cell

Assuming that the same pilot sequences are reused once in the same time-slot in all cells, then the probability to encounter pilot contamination in any cell will be $\alpha = \frac{K'}{OP}$, where OP denotes the number of orthogonal pilot sequences in the system. Following this assumption, $L' = r(L \times \alpha)$ represents the number of cells that upload the same pilot sequence, where $r(.)$ is a function that rounds to the nearest integer. Note that $L \times \alpha$ denotes the number of users per pilot, but since we assumed that every pilot is used once per cell, $L'$ represents the number of contaminated cell.

The Least Squares Estimation of the channel matrix at the j-th BS can be written as:

$$\widehat{\boldsymbol{G}}_j = \arg\min_{\boldsymbol{G}_l} \left\| \frac{1}{\sqrt{\tau P_u}} \boldsymbol{Y}_j^p - \boldsymbol{G}_l \boldsymbol{X}_j^{p\,\boldsymbol{H}} \right\|^2 \quad (5)$$

The solution of (5) can be expressed as (6):

$$\widehat{\boldsymbol{G}}_j = \sqrt{\tau P_u}\,\boldsymbol{G}_j + \sqrt{\tau P_u} \sum_{l=1, l \neq j}^{L'} \boldsymbol{G}_l + \widehat{\boldsymbol{W}}_j^{\,up} \quad (6)$$

where $\widehat{\boldsymbol{W}}_j^{\,up}$ AWGN still has i.i.d distribution with zero-mean, unit-variance and CN (0, 1).

At the forward-link, the j-th BS transmits a precoded vector to the K' UTs based on the estimated version of (6) and uses the "last estimated channel state information (CSI)" to precode the downlink vector to the K" UTs. The use of the last estimated CSI by the BS do not deteriorate the UTs' performances since the coherence interval of those K" UTs is long enough to preserve channel characteristics during several time slots. Considering the use of Eigen-beamforming linear precoder, the K×1 received vector at the K UT of the l-th cell can be represented as (7):

$$\boldsymbol{Y}_j^d = \sqrt{P_d} \sum_{l=1}^{L} \boldsymbol{G}_j^T \left[ \widehat{\boldsymbol{G}}_j + \overline{\widehat{\boldsymbol{G}}_j} \right]^* \boldsymbol{X}_j^d + \boldsymbol{W}_j^d \quad (7)$$

$\overline{\widehat{\boldsymbol{G}}_j}$ is a matrix of dimension M×K that denotes the sparse complement matrix of $\widehat{\boldsymbol{G}}_j$, where $\overline{\widehat{g}_{jmk}} \triangleq [\overline{\widehat{\boldsymbol{G}}_j}]_{m,k}$

$$\overline{\widehat{g}_{jmk}} = \begin{cases} 0 & if\ k \in SK' \\ \widehat{g}_{jmk}(t-n) & if\ k \in SK'' \end{cases}$$

SK' and SK" are the set of UTs that upload their pilots and the set of UTs that skip uploading their pilots respectively during the current time-slot.

In the same manner, $\widehat{\boldsymbol{G}}_j$ of dimension M×K denotes the sparse complement matrix of $\overline{\widehat{\boldsymbol{G}}_j}$ and

$$\widehat{g}_{jmk} = \begin{cases} \widehat{g}_{jmk}(t) & if\ k \in SK' \\ 0 & if\ k \in SK'' \end{cases}$$

where $t \in \mathbb{N}$ represent the current time-slot and $n \in \mathbb{N}$ represent the time-slot that contains the last estimated version of the channel. $\boldsymbol{X}_j^d$ is the K×1 symbols vector received by the K users in the l-th cell, $P_d$ is the normalized received SNR at each UT and the K×1 matrix $\mathbf{W}_j^{\,d}$ represents additive AWGN i.i.d noise vector with zero-mean and unit-variance.

Following the analysis in [17], as $M \gg K$ the following relation holds:

$$\left(\frac{\boldsymbol{G}_l^H \boldsymbol{G}_l}{M}\right)_{M \gg K} = \boldsymbol{D}_j^{1/2} \left(\frac{\mathbf{H}_l^H \mathbf{H}_l}{M}\right)_{M \gg K} \boldsymbol{D}_j^{1/2} \approx \boldsymbol{D}_j^{1/2} \qquad (8)$$

and $\frac{1}{M}\boldsymbol{H}_l^T \boldsymbol{H}_l^* = \boldsymbol{I}_K \delta_l$, where $\boldsymbol{I}_K$ is an Identity matrix with dimension $K \times K$ and $\delta_l$ corresponds to the covariance factor of $\mathbf{H}_l$.

The received signal at the k-th UT in the j-th cell can be deduced from (7) and (8):

$$\frac{1}{M\sqrt{\tau P_u P_d}}\mathrm{y}_{jk}^{\,d} = \beta_{jk}x_{jk}^{\,d} + \sum_{l=1,l\neq j}^{L'} \beta_{lk}x_{jk}^{\,d} \quad (9)$$

where $\beta_{lk}$ corresponds to the large-scale fading between the k-th UT in the l-th cell and the j-th BS, and $x_{jk}$ is the k-th element of symbol vector $\boldsymbol{X}_j^{\,d}$. The signal to interference noise ratio of each UT can be written as:

$$SINR = \frac{\beta_{jk}^2}{\sum_{l=1,l\neq j}^{L'} \beta_{lk}^2} \qquad (10)$$

## 3    Time-shifted Pilots

The diversity among UT coherence intervals depends on the propagation environment, user mobility, and the carrier frequency [3]. Since not all UTs encounter the same mobility within the cell, we can classify them according to their coherence interval length. Therefore, users belonging to Class 1 encounter the shortest coherence interval of length T, which is also considered as the TDD frame size. Indeed, users of Class 1 should upload their pilots each TDD frame. Furthermore, users of Class n with a coherence interval of length T'> nT should upload their pilots once each n TDD frames. Within one channel estimation during the coherence interval, BS can be precode all downlink data belong to this coherence interval based on the last estimated CSI. In Figure 1, we introduce an example of a sequence of TDD frames related to 3 UTs belonging to Class 3 using the same pilot sequence (represented by P). The 3 UTs exchange different data streams with the BS (represented by D). By shifting the frame toward a pilot free time-slot, users of Class n can reuse the same pilot n times subjected to T (n-1) ≤ Q. We denote by Q the maximum acceptable coherence interval which can be assigned by the network designers according to performance demands.

By following the definition of the sample duration of the slots, expressed in [18] as the number of OFDM symbols times, the tone duration of the Nyquist sampling interval $T = \frac{T_{slot}\,T_u}{T_s\,T_g}$. For typical OFDM parameters, symbol interval is $T_s = 1/14\ ms$, usable interval is $T_u = 1/15$ ms, guard interval is $T_g = 1/220\ ms$. Assuming delay-spread equals $T_g$, Nyquist interval is equivalent to $\frac{T_u}{T_g} = 14$ tones.

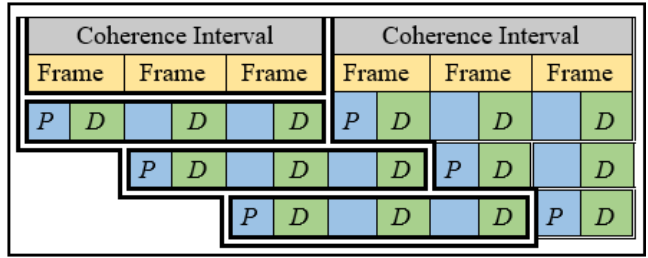

**Figure 1: Time-Shifted frames of Class 3 UTs**

Consider Class 1 corresponding to UTs moving in a train of speed 300 Km/h, thus it took 473.7µs to pass ¼ wavelength at frequency 1.9 GHz which has an equivalent sample duration of T=99. Assuming that a pedestrian uses moves with an average speed of 1.38 m/s their sample duration will be Tp= 5606 which will lead to n=60 and hence, UTs related to this class can reuse the same pilot 60 times (ignoring the maximum shift delay limit).

## 4    User Terminal Classification

In this section, we propose a UTs classification algorithm that will allow us to take benefit from the proposed pilot shifting technique. One way to classify UTs is to monitor their consecutive CSI covariance matrices and classify them according to their speed of change among several time-slots. Indeed, UTs with high mobility profiles will be assigned to lower classes and low mobility UTs will be assigned to higher classes. Assume that the BS had a prior assigned Class C(k) corresponding to the coherence interval T of UT k. The BS can run the following algorithm (Figure 2) periodically or on SNR failure, to update the assigned class C(k) of the k-th UT.

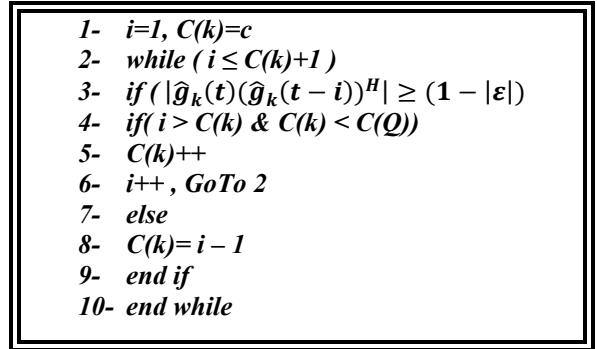

**Figure 2 User classification algorithm**

Following the algorithm, depicted in Figure 2, the BS checks the covariance of the current channel vector $\hat{g}_k(t)$ with the previous C(k) channel vectors. In the case that the channel persists with an acceptable error of $|\varepsilon|$ for C(k)+1 consecutive time-slots and still below the limit Class C(Q), the Class C(k) of the k-th UT C(k) is promoted. Otherwise, if the channel failed to persist for C(k) consecutive time-slots, the class assigned to the k-th UT will be degraded. After the k-th UT has been classified to Class C(k), he should upload his pilots once every C(k) consecutive time-slots, which will lead not only to the decrease of the uplink pilot contamination, but also to the reduction in the transmitted power at the UT and the increase of the density of UTs/cell.

Recent advances in massive MIMO systems recommend coordinating the pilot allocation between interfering cells by ensuring no pilot reuse between adjacent cells to avoid pilot contamination from the first tier of interfering cells. These techniques solve radically the pilot contamination problem for practical numbers of antennas. However, the number of UTs per cell decreases drastically since only a fraction of the pilot sequences are used in each cell. Interestingly, the proposed sparse pilot assignment allows pilot reuse within the same interference domain by shifting pilots toward a pilot free time-slot, which increases the UTs density per cell while mitigating inter and intra cell interference.

## 5    Spectral and Energy Efficiency

We follow the definition of spectral efficiency given by [2] to evaluate the spectral efficiency of our proposed model. Considering channel estimation processed using maximum ratio combination (MRC) receiver, the general uplink spectral efficiency SE, is given by:

$$SE = \frac{T-\tau}{T} K log_2(1 + SINR) \qquad (11)$$

The signal to interference noise ratio (SINR) represented as follows:

$$SINR = \frac{\tau(M-1)P_u^2}{\tau(K\bar{L'}^2 - 1 + \gamma(\bar{L'}-1)(M-2))P_u^2 + \bar{L'}(K+\tau)P_u + 1} \qquad (12)$$

where $\gamma \in [0,1]$ represents the inter-cell interference factor and $\bar{L'} \triangleq (L'-1)\gamma + 1$. The Energy Efficiency EEn for Class n is expressed as:

$$EE_n = \frac{1}{P_u}\left(\frac{n(T-\tau)}{nT-(n-1)\tau}\right) K_n log_2(1 + SINR) \qquad (13)$$

where $K_n$ is the number of users of Class n.

## 6    Numerical Results

In this section, we illustrate the performances of the proposed technique through Matlab simulations. We simulate a scenario with L=7 hexagonal cells. We assume pilot reuse once at the same time-slot in each cell, and that each cell owns τ=30 pilot sequences. The system uses a carrier frequency of 1.9 GHz and we consider UTs with several channel coherence profiles. We also assume that the transmit power P_u is upper bounded by 100 mdB. Inter-cell interference factor is γ=0.3 and K=30. We consider Class 1 time-slot of sample duration T=99 OFDM symbols and C(Q)=30.

In the first scenario, we simulate a conventional massive MIMO system (with only Class 1 users, i.e. all UTs send pilot every time slot) and a system with Class 3 users (every UTs uploads pilot once every 3-time slots). We recall that the time slot corresponds to the smaller coherence interval of all the UTs. Figures 3 and 4 illustrate respectively the energy efficiency and the spectral efficiency of Class 1 and Class 3 UTs. We can clearly observe that Class 3 overcome the classical massive MIMO in both SE and EE, and the gap between Class 1 and 3 still increase with the number of antennas. Specifically, for 100 antennas, we can see that the EE of Class 3 UTs is almost 4 times better than conventional massive MIMO system. Note that, in conventional massive MIMO system, if there is only one UT having a coherence interval of Ts (Class 1) and all the other UTs have a coherence interval higher than 3*Ts (Class 3), all the UTs have to upload pilot to the BS every Ts.

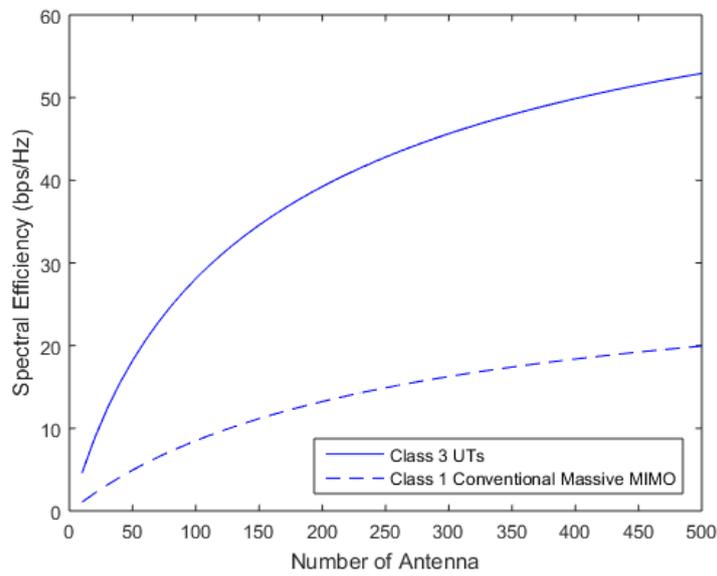

**Figure 3 Spectral Efficiency Vs M**

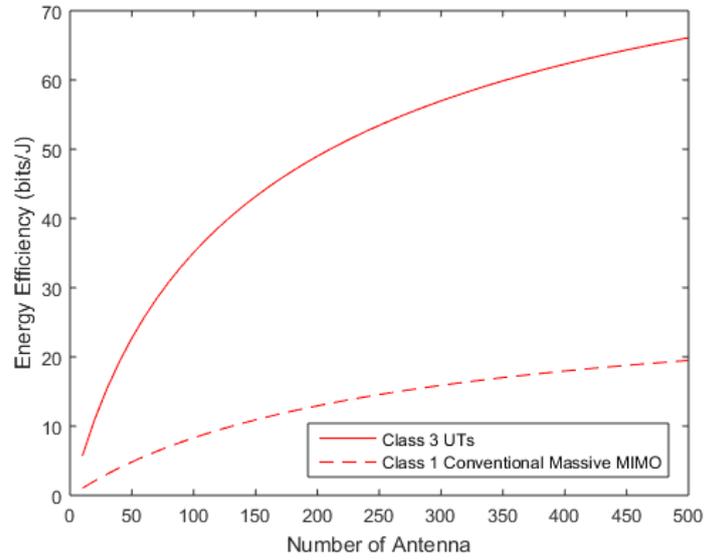

**Figure 4 Energy Efficiency Vs M**

To illustrate the performance of different classes of UTs, we further simulate the EE depending on the class index. In Figure 5, we consider M=300 antennas and vary the class index from 1 to 30, wherein Figure 6 we vary M and we illustrate the EE of the set of classes from 1 to 30. Both figures demonstrate the significant advantage of using our proposed technique in terms of both: energy efficiency and spectral efficiency.

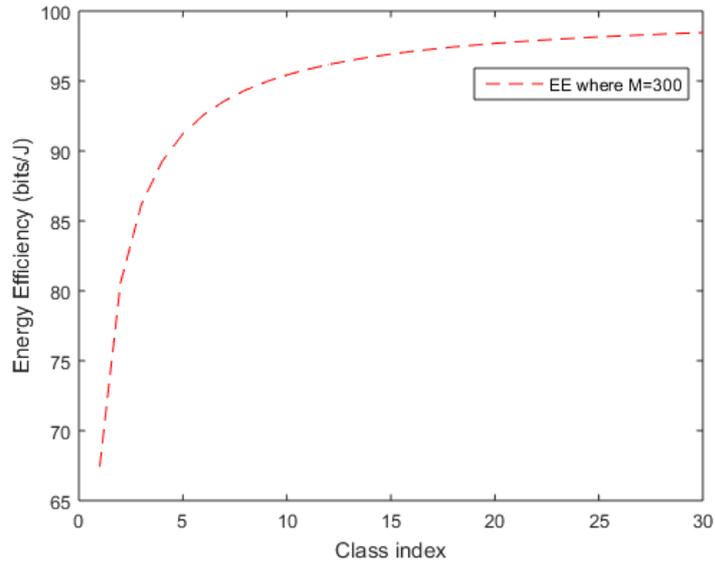

**Figure 5 Energy efficiency Vs Class index**

Indeed, there is a gap between the EE performance of the set of classes with indices greater than 1 and the EE performance of Class 1 (the curve at the bottom, Conventional massive MIMO). This result supports our claim about the utility of frame-shifting and sparse frame pilot for massive MIMO. Note also that frame-shifting and sparse frame pilot reduce the computational estimation costs at the BS due to sparse received channel matrix. Moreover, by pilot skipping, pilot reuse will lead to the increase of the number of UTs/cell without any pilot contamination. Furthermore, the transmitted power of each UT can be reduced due to the reduction of inter and intra cell interference.

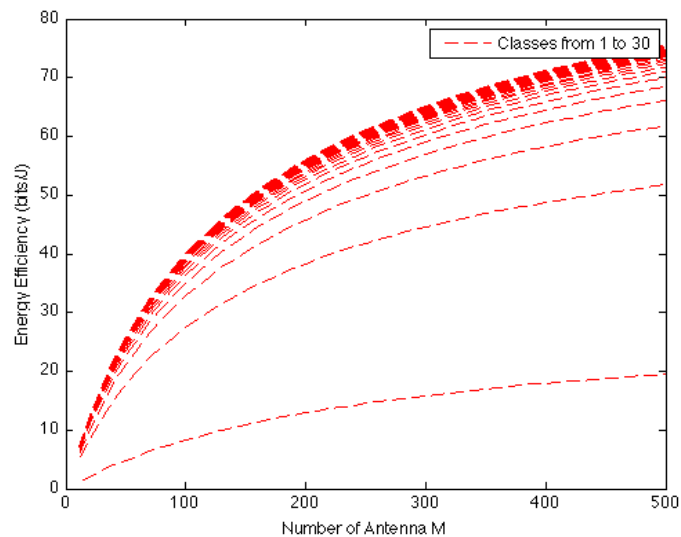

**Figure 6 Classes indices from 1 to 30 Vs EE**

## 7    Conclusion

In this paper, we have introduced a novel technique to reduce the pilot overhead by classifying UTs based on their coherence interval. Then, we have shifted frames containing pilots to an empty time pilot space. Indeed, using sparse pilot and frame-shifting, a little number of orthogonal pilot sequences are capable of serving a higher number of UTs and pilot reuse will be possible within the same cell without leading to pilot contamination. The proposed technique had proved its ability to mitigate pilot contamination, increase spectral and energy efficiency, increase UT/cell and reduces estimation computational cost. Note that channel coherence based classification among UTs should be considered as a main issue in the future 5G mobile networks.